\documentclass[aps, preprint, eqsecnum, showpacs, epsfig]{revtex4}
\usepackage{graphics}
\usepackage{graphicx}
\usepackage{epsfig}

\def \be {\begin{equation}}
\def \ee {\end{equation}}
\def \ba {\begin{eqnarray}}
\def \ea {\end{eqnarray}}
\def \bm {\begin{displaymath}}
\def \em {\end{displaymath}}
\def \br {{\bf r}}

\begin{document}
\title{Free-energy functional for freezing transitions: Hard sphere systems 
freezing into crystalline and amorphous structures}
\author{Swarn Lata Singh, Atul S. Bharadwaj and Yashwant Singh}
\author{}
\affiliation{Department of Physics, Banaras Hindu University, 
Varanasi-221 005,
India}
\date{\today}
\begin{abstract}
A free-energy functional that contains both the symmetry conserved and 
symmetry broken parts of the direct pair correlation function has been 
used to investigate the freezing of a system of hard spheres into crystalline 
and amorphous structures. The freezing parameters for fluid-crystal transition 
have been found to be in very good agreement with the results found from 
simulations. We considered amorphous structures found from the molecular 
dynamics simulations at packing fractions $\eta$ lower than the glass close 
packing fraction $\eta_{J}$ and investigated their stability compared to 
that of a homogeneous fluid. The existence of free-energy minimum 
corresponding to a density distribution of overlapping Gaussians centered 
around an amorphous lattice depicts the deeply supercooled state with a 
heterogeneous density profile. 
\end{abstract}

\maketitle
\section{\bf Introduction}
When a liquid freezes into a crystalline solid its continuous symmetry 
of translation and rotation is broken into one of the Bravais lattices. 
In three dimensions the freezing of a liquid into a crystalline solid is 
known to be first-order symmetry breaking transition marked by large 
discontinuities in entropy, density and order parameters. The molecules 
in a crystal are localized on a lattice described by a discrete set of 
vectors $\{{\bf{R}}_{i}\}$ such that any functions of position, such as one 
particle density $\rho({\br})$ satisfies $\rho(\br)=\rho(\br+{\bf{R}}_i)$ 
for all ${\bf{R}}_{i}$ $[1]$. This set of vectors necessarily forms a Bravais 
lattice. But when the liquid is supercooled bypassing its crystallization, it 
continues to remain in amorphous state. With increase in density a solid like 
phase emerges with molecules getting localized around their mean position on a 
random structure. The underlying lattice on which such localized motion takes
place is related to the time scale of relaxation in the supercooled liquid. 
While the supercooled liquid starts to attain solidlike properties, structurally
it does not have any long range order like the one present in the crystal. 
This phenomena is termed the ``glass transition'' $[2,3]$. Although the glassy 
materials are well characterized experimentally, the existence of phase 
transition into the glassy state remains controversial $[4-6]$. Our aim in 
this paper is not to enter in this discussion but to examine the stability 
(metastability) of amorphous structures from a thermodynamic point of view, 
using the standard method of density-functional theory which is also used to 
investigate the crystallization of liquids. We refer to glassy or amorphous 
solid a structure in which particles are localized around their mean positions 
forming a random lattice. Localization of particles amounts to breaking of 
continuous translational symmetry of normal liquid and takes place 
in forming both crystals and glasses.

The structural properties of a classical system can adequately 
be described by one and two-particle density distributions. 
The one particle density distribution $\rho(\br)$ defined as

\be 
\rho(\br)= \langle\sum_{k} \delta{(\br-{\bf{R}}_{k})}\rangle
\ee
where $\bf{R}_k$ is the position vector of k-th particle and bracket 
$\langle....\rangle$ represents the ensemble average, is constant independent 
of position for an isotropic liquid, but contains most of the structural 
informations of crystals and glasses. The particle density distribution 
$\rho^{(2)}(\br_{1}, \br_{2})$ which gives the probability of finding simultaneously
a molecule in a volume element $d\br_{1}$ centered at $\br_{1}$ and a second 
molecule in volume element $d\br_{2}$ centered at $\br_{2}$ is defined as

\be
\rho^{(2)}(\br_{1}, \br_{2})=\langle \sum_{j} \sum_{\neq k} \delta{(\br_{1}-\bf{R}_{j})} 
\delta{(\br_{2}-\bf{R}_{k})} \rangle
\ee

The pair correlation function $g{(\br_{1},\br_{2})}$ is related to 
$\rho^{(2)}(\br_{1}, \br_{2})$ by the relation
\be
g(\br_{1}, \br_{2})=\frac{\rho^{(2)}(\br_{1}, \br_{2})}{\rho(\br_{1}) \rho(\br_{2})}
\ee

The direct pair correlation function $c(\br_{1}, \br_{2})$ which appears 
in the expression of free-energy functional (see Sec. $II$) is related to 
the total pair correlation function $h(\br_{1}, \br_{2})=g(\br_{1}, \br_{2})-1$ 
through the Ornstein-Zernike (OZ) equation,
\be
h(\br_{1}, \br_{2})=c(\br_{1}, \br_{2})+ \int{c(\br_{1}, \br_{3}) 
\rho(\br_{3}) h(\br_{2}, \br_{3})d\br_{3}}
\ee

Since in an isotropic liquid $\rho(\br_1)=\rho(\br_2)=\rho_{l}=N/V$ 
where $N$ is the average number of molecules in volume $V$,
\be
g(r)=\frac{\rho^{(2)}(\br_{1}, \br_{2})}{\rho^{(2)}_{l}}
\ee
where $r=|\br_{2}-\br_{1}|$. In a liquid of spherically symmetric 
particles $g(\br_{1}, \br_{2})$, $h(\br_{1}, \br_{2})$, $c(\br_{1}, \br_{2})$ 
depend only on interparticle separation $r=|\br_{2}-\br_{1}|$. This 
simplification is due to homogeneity which implies continuous translational 
symmetry and isotropy which implies continuous rotational symmetry. Such 
simplification does not, in general, occur in frozen phases. We refer crystal 
as well as glass as frozen phases. While a crystal is both inhomogeneous and 
anisotropic, a glass can be regarded as isotropic but inhomogeneous. The 
heterogeneity in glassy system over length and time scales has been studied 
in several recent works $[7]$ related to computer simulations. 

The total and direct pair correlation functions of a system can be given 
as a simultaneous solution of the OZ equation and a closure relation that 
relates the correlation functions to the pair potential. Well known closure 
relations are the Percus-Yevick (PY) relation, the hypernatted chain (HNC) 
relation and the mean spherical approximation (MSA). It may, however, 
be noted that while the OZ equation $(1.4)$ is general and connects the 
total and direct pair correlation functions of liquids as well as of frozen 
phases, the closure relations have been derived assuming translational 
invariance $[8]$ and therefore valid only to normal fluids. The integral 
equation theory has been used quite successfully to describe the structure 
of isotropic liquids. But its application to frozen phases has so far been 
very limited $[9,10]$. In Sec. $III$ we describe a method to calculate the 
DPCF of frozen phases formed by breaking of continuous translational 
symmetry of liquids and use it in a free-energy functional described in 
Sec. $II$ to study freezing transitions in Sec. $IV$ and $V$. 

Since its advent in 1979 by Ramakrishnan and Yussouff (RY) $[11]$, the 
density functional theory (DFT) has been applied to freezing transition 
of variety of pure liquids and mixtures $[11,12]$. A DFT requires an 
expression of the Halmholtz free-energy (or grand thermodynamic potential) 
in terms of one- and two-particle distribution functions and a relation that 
relates $\rho(\br)$ to pair correlation functions. Such a relation is formed 
by minimizing the free-energy with respect to $\rho(\br)$ with appropriate 
constraints $[12,13]$. The DPCF that appear in these equations are of frozen 
phase and are functional of $\rho(\br)$. When this functional dependence is 
ignored and the DPCF is replaced by that of the co-existing isotropic liquid 
$[11]$ or by an ``effective fluid'' $[14]$ the free-energy functional becomes 
approximate and fails to provide an accurate description of freezing transitions 
and stability of frozen phases. An improved free-energy functional which contains 
both the symmetry conserved and symmetry broken parts of the DPCF has recently 
been developed $[9,10]$ and applied to study the isotropic-nematic transition 
$[9]$ and the crystallization of power-law fluids $[10]$. 

In this paper we investigate the freezing of fluids of hard sphere 
into crystalline and amorphous phases and compare our results with 
the results of previous investigations. Hard spheres are ubiquitous in 
condensed matter; they have been used as models of liquids, crystals, 
colloidal systems, granular systems and powders. Packing of hard spheres 
are of even wider interest as they are related to important problems in 
information theory, such as diagonalization of signals, error correcting 
codes, and optimization problems $[15]$. Recently, amorphous packing of 
hard spheres have attracted much attention $[6,16-18]$ because for 
polydisperse colloids and granular materials the crystalline state is 
not obtained for kinetic reasons . It is therefore necessary to have a 
statistical-mechanical theory based on first principle which can 
correctly describe the freezing of hard spheres.

The paper is organized as follows: In Sec. $II$ we describe the 
free-energy functional for a symmetry broken phase which contains 
both the symmetry conserving and symmetry broken parts of direct pair 
correlation functions. In Sec. $III$ we describe a method to calculate 
these correlation functions. The theory is applied in Sec. $IV$ to 
investigate the freezing of fluids into crystalline solids and in Sec. 
$V$ the metastability of amorphous structures.

\section{ Free-energy functional} 

An important step in construction of the density functional 
model of a frozen phase is proper parametrization of extremely 
inhomogeneous density function $\rho(\br)$ which value near a lattice 
site may be orders of magnitude higher than in the interstitial 
regions. One very successful prescription of $\rho(\br)$ is as a collection 
of overlapping Gaussian profiles $[19]$ centered over a set of lattice 
sites ${\bf{R}}_{m}$.
\be
\rho(\br)=\sum_{m} (\frac{\alpha}{\pi})^{3/2} exp\left[-\alpha
{\left(\br-{\bf{R}}_{m}\right)}^{2}\right]
\ee

The width of a Gaussian profile is inversely proportional to the square 
root of $\alpha$ which will be referred to as localization parameter. 
In this representation, the limit $\alpha \rightarrow 0$ is the 
homogeneous liquid state and higher values of $\alpha$ represent 
increasingly localized structures.

Taking Fourier transform of $(2.1)$ one gets
\be
\rho(\br)=\rho_{0}+\frac{1}{V}\sum_{q\neq 0} \rho_{q} 
e^{i\bf {q}\cdot \br}
\ee
where 
\be
\rho_{q}=e^{-{q^{2}}/4\alpha} \sum_{n}e^{-i \bf {q} \cdot {\bf{R}}_{n}}
\ee
is amplitude of density wave of wavelength $2\pi/|q|$. The nature 
and magnitude of inhomogeneity of a frozen phase is measured by 
$\rho_q$ which will be referred to as order parameter; $\rho_{q}=0$ 
for $q\neq 0$ corresponds to isotropic fluid and $\rho_{q}\neq 0$ to 
a frozen phase. For a crystal in which ${{\bf{R}}_{m}}$ forms a periodic 
lattice, $e^{i \bf{q} \cdot \bf{R}_{m}}=\delta_{\bf{q}, \bf{G}}$ 
where $\bf{G}$ are reciprocal lattice vectors (R. L. V.), Eq.$(2.2)$ 
reduces to 
\be
\rho_{s}(\br)=\rho_{0}+\rho_{0} \sum_{G} e^{-G^{2}/4\alpha} 
e^{i \bf{G} \cdot \br}
\ee
   
This is a well known expression of $\rho (\br)$ of a crystal. As in an 
amorphous or glassy structure the lattice sites are randomly distributed 
and are not known, the above simplification is not possible. We can, 
however, calculate and have an idea of inhomogeneity and its difference 
from that of a crystalline solid using the distribution of $\bf{R}_{n}$
determined from numerical simulations $[20,21]$. Thus
\be
\rho_{g}(r)=\rho_{0}+\frac{1}{V}\sum_{q\neq 0} e^{-{q^{2}}/{4\alpha}} 
\left [1+\rho_{0}
\left(\int d{\bf{R}} h(R)e^{-i{\bf{q}}\cdot {\bf{R}}}\right)\right] 
e^{i {\bf{q}} \cdot {\bf{r}}}
\ee
where $g(R)=1+h(R)$ is the site-site correlation function which 
provides the structural description of amorphous structure.
In Fig.$1$ we plot $\rho_{g}(\br)$ for $\eta(\equiv \frac{1}{2}\pi \rho_{0} 
{\sigma}^{3}; \sigma$ being the diameter particle)$=0.576$ and $\alpha=150$ 
(highly localized condition) and $\alpha=15$ (weakly localized condition). 
In calculating $\rho_{g}(r)$ we used $g(R)$ data found for amorphous 
structures of granular particles [21] shown in Fig.$2$. In Fig.$2$ we also 
show the value of $g(R)$ of a liquid found from solving the integral 
equations discussed in Sec. $III$, for comparison. From Fig.$1$ we see 
that while inhomogeneity increases with value of $\alpha$, unlike crystalline 
solid, it remains confined to about $4$ particle diameter, even for 
$\alpha=150$ and decays rapidly on increasing the distance.

The reduced free-energy $A[\rho]$ of an inhomogeneous system is a functional 
of $\rho(\br)$ and is written as

\be
A[\rho]=A_{id}[\rho]+A_{ex}[\rho]
\ee
The ideal gas part is exactly known and is given as
\be
A_{id}[\rho]=\int d\br \rho(\br)\left[ln\left(\rho(\br)\Lambda\right)-1\right]
\ee

where $\Lambda$ is cube of the thermal wavelength associated with a molecule. 
The excess part arising due to intermolecular interactions is related to the 
DPCF as $[9,10]$
\be
\frac{\delta^{2}A_{ex}[\rho]}{\delta \rho(\br_1) \delta \rho(\br_2)}=
-c^{(0)}(\br_{1},\br_{2};\rho_{0})-c^{(b)}(\br_{1},\br_{2};[\rho])
\ee
where superscripts $(0)$ and $(b)$ represent, respectively, the symmetry 
conserving and symmetry breaking parts of the DPCF. In other words $c^{(0)}$ 
is found by treating the system to be isotropic and homogeneous with  density 
$\rho_{0}$ whereas $c^{(b)}$ are the contribution which arise due to
heterogeneity in density in a frozen phase.

$A_{ex}[\rho]$ is found by functional integration of Eq.$(2.8)$. In this 
integration the system is taken from some initial density to the final 
density $\rho(\br)$ along a path in the density space; the result is 
independent of the path of integration. As the symmetry conserving part $c^{(0)}$ 
is function of density the integration in the density space is done taking an 
isotropic fluid of density $\rho_{0}$ (or $\rho_{l}$, the density of co-existing 
fluid in case of crystal) as reference.This leads to
\be
A_{ex}^{(0)}[\rho]=A_{ex}(\rho_{0})-\frac{1}{2}\int d\br_{1} \int d\br_{2} 
\Delta \rho(\br_{1})\Delta \rho(\br_{2})c^{(0)}(r)
\ee
where $\Delta \rho(\br_{1})=\rho(\br_{1})-\rho_{0}$ and $A_{ex}(\rho_{0})$ 
is the excess reduced free-energy of an isotropic system of density 
$\rho_{0}$.

The integration over $c^{(b)}$ has to be done in the density space 
spanned by the number density $\rho_{0}$ and order parameters $\rho_{q}$. 
We characterize this part of the density space by two parameters $\lambda$ 
and $\xi$ which vary from $0$ to $1$. The parameter $\lambda$ raises the 
density from $0$ to $\rho_{0}$ as it varies from $0$ to $1$ whereas the 
parameter $\xi$ raises the order parameters from $0$ to $\rho_{q}$ for each 
value of q. This gives $[9,10]$. 
\be
A^{(b)}_{ex}[\rho]=-\frac{1}{2}\int d\br_{1} \int d\br_{2} \Delta \rho(\br_{1}) 
\Delta \rho(\br_{2}) \bar{c}^{(b)}\left(\br_{1},\br_{2};[\rho]\right)
\ee
where
\be
\bar{c}^{(b)}(\br_{1},\br_{2};[\rho])=4\int_{0}^{1} d\xi \xi \int_{0}^{1} 
d\xi^{'} \int_{0}^{1} d\lambda  \lambda \int_{0}^{1} d\lambda^{'} c^{(b)}
(\br_{1},\br_{2};\lambda \lambda^{'}\rho_{0};\xi \xi^{'}\rho_{q})
\ee

While integrating over $\lambda$ the order parameter $\rho_{q}$ are kept 
fixed and while integrating over $\xi$ the density is kept fixed. The 
result does not depend on the order of integration. The free-energy functional 
of a frozen phase is the sum of 
$A_{id}$, $\Delta A^{(0)}[\rho]$ and $\Delta A^{(b)}[\rho]$
given, respectively, by Eq.$(2.7)$,$(2.9)$ and $(2.10)$.

The minimization of $\Delta A=A[\rho]-A(\rho_{0})$ where $A(\rho_{0})$ 
is the free-energy of an homogeneous and isotropic system of density 
$\rho_{0}$, leads to
\be
\ln \frac{\rho(\br)}{\rho_{0}}=\lambda + \int d\br_{2} \Delta 
\rho(\br_{2}) c^{(0)}(|\br_{2}-\br_{1}|,\rho_{0})+\int d\br_{2} 
\Delta \rho(\br_{2}) \tilde{c}^{(b)}(\br_{1},\br_{2}) 
\ee

Here $\lambda$ is the Lagrange multiplier and is determined from the condition 
\be
\frac{1}{V}\int_{V} \frac{\rho(r)}{\rho_{0}} d\br=1
\ee
and
\be
\tilde{c}^{(b)}(\br_{1},\br_{2})=2\int_{0}^{1} d\lambda \int_{0}^{1} 
d\xi c^{(b)}(\br_{1},\br_{2},\lambda \rho_{0},\xi\rho_{q})
\ee

In principle, the only information we need to know is the value of 
$c^{(0)}(r)$ and $c^{(b)}(\br_{1},\br_{2};[\rho])$ to calculate self 
consistently the value of $\rho(r)$ which minimizes the free-energy. 
In practice, one, however, finds it convenient to do minimization for 
assumed form of $\rho(\br)$ $[12]$

\section{Calculation of Direct Pair Correlation Function}

To calculate the values of $c^{(0)}(r)$ we use the integral 
equation theory consisting of OZ equation,
\be
h^{(0)}(r)=c^{(0)}(r)+\rho_{0}\int d\br^{'} c^{(0)}(r^{'})h^{(0)}
(|\br^{'}-\br|)
\ee
and a closure relation proposed by Rogers and Young [22].
This closure relation is written as
\be
1+h^{(0)}(r)=exp(-u(r)/k_{B}T)\left[ 1+\frac{exp(\gamma(r)f(r))-1}
{f(r)}\right] 
\ee
where
\be
\gamma(r)=h(r)-c(r)
\ee
and
\be
f(r)=1-exp(-\psi r)
\ee

Here $\psi$ is an adjustable parameter used to achieve thermodynamic 
consistency and its value for a system of hard spheres is found to be 
equal to $0.16$ $[22]$. In Eq.$(3.2)$ $u(r)$ is pair potential, $k_{B}$, 
the Boltzmann constant and T, temperature. This closure relation was 
found by mixing the PY and HNC closure relations in such a way that at 
$r=0$ it reduces to the PY approximation and for values of $r$ where $f(r)$ 
approaches $1$, it reduces to the HNC approximation. Eqs$(3.1)-(3.4)$
together constitute a thermodynamically consistent theory and has been 
found to give values of pair correlation functions which are in very 
good agreement with Monte Carlo results.

The differentiation of Eqs.$(3.1)$ and $(3.2)$ with respect to density 
yields the following two relations,
\be
\frac{\partial h^{(0)(r)}}{\partial \rho_{0}}=\frac{\partial c^{(0)}(r)}
{\partial \rho_{0}}+\int d\br^{'} c^{(0)}(r^{'}) h^{0}(|\br^{'}-\br|)+
\rho_{0}\int d\br^{'} \frac{\partial c^{(0)}}{\partial \rho_{0}} h^{(0)}
(|\br^{'}-\br|)+\rho_{0}\int d\br^{'}c^{(0)}(r)
\frac{\partial h^{(0)}(|\br^{'}-\br|)}{\partial \rho_{0}}
\ee
and
\be
\frac{\partial h^{(0)}(r)}{\partial \rho_{0}}= exp\left(-\frac{u(r)}{k_{B}T} 
\right)\left[exp\left[\gamma(r)f(r)\right] \frac{\partial \gamma(r)}
{\partial \rho_{0}} \right]
\ee

The closed set of coupled equations $(3.1)$, $(3.2)$, $(3.5)$ and 
$(3.6)$ have been solved using Gillen's algorithm $[23]$ for four unknowns 
$h^{(0)}$, $c^{(0)}$, $\frac{\partial h^{(0)}}{\partial \rho_{0}}$ and 
$\frac{\partial c^{(0)}}{\partial \rho_{0}}$. We compare the values of 
$c^{(0)}(r)$ at packing fractions $\eta=0.50$ and $0.55$ in Fig.$(3)$ and 
values of $\frac{\partial c^{(0)}(r)}{\partial \rho_{0}}$ for the same 
two values of $\eta$ in Fig.$(4)$ to see the density dependence of these 
functions. While $\eta=0.50$ is close to the value of packing fraction 
at which system freezes into a crystalline solid, $\eta=0.55$ is close 
to the value of packing fraction $\eta_{M}$ at which the crystal melts. 

Since all closure relations (including the one given by Eq.$(3.2)$) 
which are used in the integral equation theory for pair correlation 
functions are derived assuming translational invariance $[8]$, their use 
in calculating the values of pair correlation function of frozen phases may 
not be appropriate. In view of this we use a series expansion in which the 
contribution to the the DPCF arising due to inhomogeneity of the system 
is expressed in terms of higher body direct correlation functions
of the uniform (isotropic and homogeneous) system. Thus
\begin{equation}
 c^{(b)}(\br_{1},\br_{2};[\rho])=\int d\br_{3}  
c^{(0)}_{3}(\br_{1}, \br_{2}, \br_{3}; \rho_{0})(\rho(\br_{3})-\rho_{0})
+\int d\br_{3} d\br_{4} c^{(0)}_{4}(\br_{1}, \br_{2}, \br_{3}, \br_{4}) 
(\rho(\br_{3})-\rho_{0})(\rho(\br_{4})-\rho_{0})+\cdots
\end{equation}
where $\rho(\br_{n})-\rho_{0}=\sum_{q\neq 0}\rho_{q}e^{i\bf{q}\cdot \br_{n}}$. 
In Eq.$(3.7)$ $c^{(0)}_{n}$ are n-body direct correlation functions of 
the uniform system which can be found using the relations
\begin{equation}
\frac {\delta c^{(0)}(r)}{\delta \rho_{0}}=\int d\br_{3} c^{(0)}_{3} 
(\br_{1}, \br_{2}, \br_{3})
\end{equation}
\begin{equation}
\frac{\delta^{2}c^{(0)}(r)}{\delta \rho_{0}^{2}}=\int d\br_{3} d\br_{4}
c^{(0)}_{4}(\br_{1}, \br_{2}, \br_{3}, \br_{4})
\end{equation}

etc. The values of derivatives of $c^{(0)}(r)$ appearing on l. h. s. 
of above equations have been found using the integral equation 
theory described above.

We note that Eq.$(3.7)$ satisfies the condition that $c^{(b)}$ is zero 
in the fluid phase and depends on the magnitude (order parameter) and 
phase factors of density waves. These density waves measure the nature and
magnitude of inhomogeneity of frozen phases. While each wave contributes 
independently to the first term of Eq.$(3.7)$ interactions between the 
two waves contribute to the second term and so on. The contributions 
made by successive terms of Eq.$(3.7)$ depends on the range of pair 
potential $u(r)$ $[10]$. As the range of potential increases the 
contribution made by higher order terms increases. For a system of 
hard spheres we find that at the freezing transition the contributions 
made by first term to free-energy is already small and therefore higher 
terms are expected to be negligible; it is only for $u(r)\propto r^{-n}, n<12$ 
the contribution made by second order term becomes important $[10]$. 
In view of fast convergence of the series, Eq.$(3.7)$ seems 
to be a useful expression for calculating $c^{(b)}(\br_{1}, \br_{2})$ 

The first term of Eq.$(3.7)$ involves three-body direct correlation
function which can be factorized as a product of two-body functions $[24]$. 
Thus
\begin{equation}
c^{(0)}_{3}(\br_{1}, \br_{2}, \br_{3}; \rho_{0})=t(r_{12})t(r_{13})t(r_{23}) 
\end{equation}
The function $t(r)$ is determined using relation of Eq. $(3.8)$

\begin{equation}
\frac{\partial c^{(0)}(r)}{\partial \rho_{0}}=t(r)\int d\br^{'} t(r^{'})
t(|\br^{'}-\br|)
\end{equation}

We adopt the numerical procedure developed in [24] to calculate $t(r)$
from known values of $\frac{\delta c^{(0)}(r)}{\delta \rho_{0}}$ from
$(3.11)$. The values of $t(r)$ are plotted in Fig.$(5)$ for 
$\eta=0.50$ and $0.55$ to show their density dependence.

Taking only the first term in Eq.$(3.7)$ we write,
\begin{equation}
 c^{(b)}(\br_{1}, \br_{2})=\frac{1}{V} \sum_{q} \sum_{n} \mu_{q} 
\int d\br_{3} t(|\br_{3}-\br_{1}|) e^{i \bf{q}(\br_{3}-\bf{R_{n}})} 
t(|\br_{3}-\br_{2}|)
\end{equation}
where $\mu_{q}=e^{-q^{2}/4\alpha}$
Using the relation
\begin{equation}
\br_{3}= \frac{1}{2}(\br_{1}+\br_{2})-\frac{1}{2}(\br_{2}-\br_{1})+
(\br_{3}-\br_{1})
\end{equation}

we find that $c^{(b)}(\br_{1}, \br_{2})$ can be written in a Fourier series
in the center of mass variable $\br_c=\frac{1}{2}|\br_{1}+\br_{2}|$ with
coefficients that are functions of the difference variable 
$\br= \br_{2}-\br_{1}$, i.e,

\begin{equation}
c^{(b)}(\br_{1}, \br_{2})=\frac{1}{V} \sum_{q} c^{(q)}(\br) 
e^{i \bf{q} \cdot \br_{c}}
\end{equation}
where
\begin{equation}
c^{(q)}(\br)=\sum_{n} \mu_{q} e^{-i \bf{q} \cdot \bf{R_{n}}}
e^{-\frac{1}{2} \bf{q} \cdot \br} \int d\br^{'} t(r^{'}) 
e^{i \bf{q} \cdot \br^{'}} t(|\br^{'}-\br|)
\end{equation}

Since the DPCF is real and symmetric with respect to the interchange 
of $\br_{1}$ and $\br_{2}$, $c^{(q)}(\br)=c^{(-q)}(\br)$ and 
$c^{(q)}(\br)=c^{(q)}(-\br)$. For given value of $\alpha$ and 
${\bf{R_{n}}}$ one can calculate $c^{(q)}(r)$ from known value of $t(r)$. 
We discuss our results for crystalline and amorphous solids in 
following sections.

\section{Crystalline Solid}

{\bf{A. \underline{Calculation of $c^{(b)}(\br_{1}, \br_{2})$}}}

For a crystal in which vectors ${\bf{R}_{n}}$ form a Bravais lattice
$Eq.(3.14)$ and $(3.15)$ can be written as $[10]$
\begin{equation}
 c^{(b)}(\br_{1}, \br_{2})=\sum_{G} c^{(G)}(\br) e^{i \bf{G} \cdot \br_{c}}
\end{equation}
and
\begin{equation}
 c^{(G)}(\br)=\rho_{0} \mu_{G} e^{-\frac{1}{2}i \bf{G} \cdot \br} 
\int d\br^{'} t(r^{'}) e^{i \bf{G} \cdot \br^{'}} t(|\br^{'}-\br|)
\end{equation}
where $\mu_{G}=e^{-G^{2}/4\alpha}$
$Eq.(4.2)$ has been solved using Rayleigh expansion to give
\begin{equation}
 c^{(G)}(\br)=\sum_{lm} c^{(G)}_{lm}(r)Y_{lm}(\hat r)
\end{equation}
where
\begin{equation}
 c^{(G)}_{lm}(r)=\frac{\rho_{0}\mu_{G}}{2\pi^{2}} \sum_{l_{1}}\sum_{l_{2}}
(i)^{l_{1}+l_{2}}(-1)^{l_{2}}\left [\frac{(2l_{1}+1)(2l_{2}+1)}{2l+1} \right ] 
[C_{g}(l_{1}, l_{2}, l;0, 0, 0)]^{2} j_{l_{2}}(\frac{1}{2}Gr)t(r)
B_{l_{1}}(r,G) Y^{*}_{lm}(\hat G)
\end{equation}
Here $C_{g}$ is the Clebsch-Gordan coefficient, $j_{l}(x)$ the spherical 
Bessel function and
\begin{equation}
 B_{l_{1}}(r,G)=(4\pi)^{2} \int dk k^{2} t(k) j_{l_{1}}(kr)\int dr^{'} r^{'^{2}}
t(r^{'}) j_{l_{1}}(kr^{'}) j_{l_{1}}(Gr^{'})
\end{equation}

The crystal symmetry dictates that $l$ and $l_{1}+l_{2}$ are even and 
for a cubic crystal, $m=0,\pm 4$. The $c_{lm}^{(G)}(r)$ depends on 
the order parameter and on the magnitude of R. L. V.

The Fourier transform of $c^{(G)}(\br)$ defined as 
\begin{equation}
 c^{(G)}({\bf{k}})=\int c^{(G)}(\br) e^{-i {\bf{k}} \cdot \br} d\br
=\sum_{lm}c^{(G)}_{lm}(k) Y_{lm}(\hat{k})
\end{equation}
where 
\begin{equation}
 c^{(G)}_{lm}(k)=4\pi(i)^{l} \int dr r^{2} j_{l}(kr) c^{(G)}_{lm}(r)
\end{equation}
is calculated from the knowledge of $c^{(G)}_{lm}(r)$. In Figs$(6)$ we plot 
$c^{(G)}_{lm}(k)$ for the first two sets of G at $\eta=0.55$ and $\alpha=170$ 
for face centered cubic (f.c.c.) structure.

{\bf{B. \underline{Liquid-Solid transition}}}

The grand thermodynamic potential defined as $-W=A-\beta \mu \int d\br \rho(\br)$, 
where $\mu$ is the chemical potential, is used to locate transition as it ensures 
that the pressure and chemical potentials of two phases remain equal at the 
transition. The transition point is determined by the condition $\Delta W= W_{l}-W=0$ 
where $W_{l}$ is the grand thermodynamic potential of the fluid. Using 
expressions given in Sec. $II$ we find
\begin{equation}
 \frac{\Delta W}{N}= \frac{\Delta W_{id}}{N}+ \frac{\Delta W_{0}}{N}+ 
\frac{\Delta W_{b}}{N}
\end{equation}

where

\begin{equation}
\frac{\Delta W_{id}}{N}=1-\ln \rho_{l}+(1+\Delta \rho^{*}) 
\left [\frac{3}{2} \ln (\frac{\alpha}{\pi})-\frac{5}{2} \right ]
\end{equation}

\begin{equation}
 \frac {\Delta W}{N}=-\frac{1}{2} \Delta {\rho^{*}}^{2} \hat {c}^{(0)}(0)- 
\frac{1}{2}\sum_{G\neq 0}(1+\Delta \rho^{*})^{2} |\mu_{G}|^{2} \hat c^{(0)}({\bf{G}})
\end{equation}
\begin{equation}
 \frac{\Delta W_{b}}{N}=-\frac{1}{2}\sum_{G_{1}}{'}\sum_{G}{'}(1+\Delta \rho{*})^{2} 
\mu_{G} \mu_{-G-G_{1}} \hat{\bar{c}}^{(G)}({\bf{G}}_1+\frac{1}{2} {\bf{G}})
\end{equation}

Here $\Delta W_{id}$, $\Delta W_{0}$ and $\Delta W_{b}$ are, respectively, the ideal,
symmetry conserving and symmetry broken contribution to $\Delta W$, the prime on 
summations in Eq.$(4.11)$ indicates the condition, $\bf{G}\ne0$, $\bf{G}_{1}\ne 0$
and $\bf{G}+\bf{G}_{1}\ne 0$, and

\begin{equation}
\hat c^{(0)}({\bf{G}})=\rho_{l} \int c^{(0)}(r) e^{i {\bf{G}} \cdot \br} d\br
\end{equation}
\begin{equation}
\hat{\bar{c}}({\bf{G}}_{1} +\frac{1}{2} {\bf{G}})=\rho_{0} \sum_{lm} \int 
\bar c^{G}_{lm}(r, \rho_{0}) e^{i({\bf{G}}_{1}+\frac{1}{2} {\bf{G}}) \cdot \br} 
Y_{lm}(\hat r) d\br 
\end{equation}
where $\rho_{0}=\rho_{l}(1+\Delta \rho^{*})$

We used the above expression to locate the fluid-f.c.c. solid and fluid-b.c.c. 
solid transitions by varying the values of $\rho_{l}$, $\Delta \rho^{*}$ and $\alpha$.
While we find fluid-f.c.c. solid transition to take place at $\eta_{l}=0.490$, 
$\Delta \eta^{*}=0.106$ and $\alpha=170$, no transition is found for b.c.c. solid. 
In table $1$ we compare our results of freezing parameters with those found by
Monte Carlo simulation [25,26] and from other density functional schemes 
[27-30]. The agreement found between our results and those of simulations 
are very good, better than any other density functional schemes.

At the transition point the contribution of different terms of Eq.$(4.8)$ is
 as follows; $\frac{\Delta W_{id}}{N}=4.44$, $\frac{\Delta W_{0}}{N}=-4.10$, 
$\frac{\Delta W_{b}}{N}=-0.34$. The contribution made by symmetry breaking term 
of free-energy is about $8.3\%$ to that of the symmetry conserving term. 
This is in accordance with the result found earlier $[10]$ for inverse power 
potential $u(r)=\epsilon (\sigma/r)^{n}$, where $\epsilon$, $\sigma$ and $n$ 
are potential parameters, that as n increases($n=\infty$ corresponds to hard 
sphere potential) the contribution made by the symmetry breaking term to 
free-energy decreases. This explains why RY theory $[11]$ while gave good 
results for hard spheres system, failed for potentials which have soft 
repulsion and/or attractive tails.

\section{Amorphous Structure}

In this section we investigate the heterogeneous density profile of an amorphous 
structure and examine the question of having metastable states in between the 
normal fluid state and the regular crystalline state at packing fraction $\eta$ 
which lies between packing fraction at melting point of a crystal $\eta_{M}=0.545$ 
and packing fraction corresponding to ``glass close packing'', $\eta_{J}\simeq=0.65$.
The usual way to construct amorphous structure in experiment or numerical simulation 
is to compress the system according to some protocol which avoids crystallization 
$[17,20,21]$. One of the criteria used to signal the onset of glassy phase in
supercooled liquids is emergence of split second peak. There may be infinitely 
large number of such metastable structures which when compressed jam along a continuous 
finite range densities down to the glass close packing $\eta_{J}$ $[31,32]$.

The density functional approach provides the means to test if such a structure 
is stable compared to that of a fluid at a given temperature and density. In 
earlier calculations the random closed packed structure generated through 
Bennett's algorithm $[33]$ was used. The $g(R)$ giving the distribution of 
particles at a given value of $\eta$ was found using an ad hoc scaling relation $[34]$.
\begin{equation}
g(R)=g_{B}[R (\frac{\eta}{\eta_{J}})^{1/3}]
\end{equation}
where $\eta_{J}$ was used as a scaling parameter such that at $\eta=\eta_{J}$ the 
random closed packed structure $g_{B}(R)$ was obtained. While Singh et. al. $[35]$ found 
that the state corresponding to this structure becomes more stable than fluid for 
$\eta\geq 0.59$ for very large value of $\alpha$ $(\sim 280)$ that corresponds to 
highly inhomogeneous density distribution, Kaur and Das $[36]$ found 
that the same structure also becomes more stable than fluid for $\eta\geq 0.576$ 
for considerably smaller values of $\alpha$ $(\simeq 18)$. On the other hand, 
Dasgupta $[37]$ has numerically located the ``glassy'' minimum of a free-energy 
functional and the structure which gave this minimum. Here we use the value of $g(R)$ 
found for granular particles from molecular dynamics simulations $[21]$ at $\eta=0.576$ 
and $0.596$ and examine the stability of these structures. The reason for choosing 
these data is that they are available at $\eta< \eta_{J}$ and have the essential 
features of an amorphous structure.

\vspace{0.5cm}
{\bf{A. \underline{Calculation of $c^{(b)}(\br_{1}, \br_{2})$}}}

From the known values of $g(R)$ the order parameter defined by Eq.$(2.3)$ is 
calculated. Thus, for $q\neq 0$
\begin{equation}
 \rho_{q}=\sum_{n} e^{-q^{2}/4\alpha} e^{-i{\bf{q}}\cdot {\bf{R}_{n}}}= 
\mu_{q} S_{a}(q)
\end{equation}
where $\mu_{q}=e^{-q^{2}/4\alpha}$ and $S_{q}(a)=1+24\eta \int dR R^{2} \left(g(R)-1\right) 
j_{0}(qR)$. In Fig.$(7)$ we plot $\rho_{q}$ as a function of $q$ for $\alpha=15$ 
and $50$ and $\eta=0.596$. The value of order parameter $\mu_{G}$ of a f.c.c 
crystal for $\alpha=50$ and $\eta=0.596$ are also shown for comparison. From 
the figure one may note that the value of $\rho_{q}$ of an amorphous structure 
has a very different magnitude and dependence on wave vector ${\bf{q}}$ than 
that of a crystal.

Using the fact that an amorphous structure can be considered on the average 
to be isotropic, Eq.$(3.15)$ is simplified to give
\begin{equation}
 c^{(q)}(r)=\frac{1}{8\pi^{3}} \mu_{q} S_{a}(q) \sum_{q}(2l+1)
B_{l}(q,r)j_{l}(\frac{1}{2}qr)
\end{equation}

In Fig.$(8)$ we show the value of $c^{(q)}(r)$ as a function of $r$ for
$q$ at which $\mu_{q}S_{a}(q)$ is maximum for $\alpha=15$ and $\eta=0.596$.

{\bf{B. \underline{Determination of free-energy minimum}}}

We calculate the minimum of $\Delta A[\rho]=A[\rho]-A_{l}[\rho_{l}]$
where $A_{l}(\rho_{l})$ is the reduced free-energy of an isotropic 
fluid of density $\rho_{0}$ and $A[\rho]$ is the reduced free-energy
of an amorphous structure of average density $\rho_{0}$. 
Using expression given in Sec. $II$ we get 
\begin{equation}
\frac{\Delta A[\rho]}{N}=\frac{\Delta A_{id}[\rho]}{N}+
\frac{\Delta A_{0}[\rho]}{N}+\frac{\Delta A_{b}[N]}{N} 
\end{equation}

\begin{equation}
\frac{\Delta A_{id}}{N}=4\pi({\frac{\alpha}{\pi})}^{3/2}\int dr r^{2} 
e^{-\alpha r^{2}} \ln [\{{(\frac{\alpha}{\pi})}^{3/2} e^{-\alpha r^{2}}\}+ 
\newline
\frac{\rho_{0}}{r}{(\frac{\alpha}{\pi})}^{1/2}
\int dRRg(R)\{e^{-\alpha {(r-R)}^{2}}-e^{-\alpha {(r+R)}^{2}}\}]
\end{equation}
\hspace{12cm} for $\alpha<20$

\begin{equation}
\frac{\Delta A_{id}}{N}=1-\ln \rho_{0}+\frac{3}{2} 
[\ln (\frac{\alpha}{\pi})-\frac{5}{3}] \hspace{2cm} 
\end{equation}
\hspace{12cm} for $\alpha>20$

\begin{equation}
\frac{\Delta A_{0}[\rho]}{N}=-\frac{1}{2}\sum_{q}|\mu_{q}|^{2} 
S_{a}(q) \hat c^{(0)}(q) 
\end{equation}
\begin{equation}
\frac{\Delta A_{b}[\rho]}{N}=-\frac{1}{2}\sum_{q}\sum_{q_{1}} 
\mu_{q_{1}}\mu_{-q-q_{1}} S_{a}(q_{1})\hat{\bar{c}}^{(q)}(|{\bf{q}}_{1}+
\frac{1}{2}{\bf{q}}|) 
\end{equation}
where
\ba
\hat{\bar{c}}^{(q)}({\bf{q}}_{1}+{\frac{1}{2}}{\bf{q}})=\int {\bar{c}}^{(q)}(r)
e^{i(\bf{q_{1}}+\frac{1}{2}{\bf{q}})\cdot \br} d\br  \nonumber 
\ea
and 
\ba
\bar{c}^{(q)}(r)=4\int d\lambda \lambda \int d\lambda^{'} d\xi \xi \int d\xi^{'}\nonumber
c^{(q)}(r, \lambda \lambda^{'}\rho_{0}, \xi \xi^{'}\rho_{q}) \\ \nonumber
\ea
As shown in Fig$(9)$, a minimum of free-energy is found at 
$\alpha\simeq 7$ for both  $\eta=0.576$ and at $\eta=0.596$.
The width of a Gaussian density profile $1/\sqrt{\alpha}\simeq 0.37$.
This is a case of weak localization. From the figure it is clear that the 
``glassy minimum'' is separated from the liquid minimum by a barrier located 
at $\alpha \simeq 0.6$ which height grows with the density. The free-energy 
minimum corresponds to a density distribution of overlapping Gaussians 
centered around an amorphous lattice and depicts the deeply supercooled
state with heterogeneous density profile.  

The schematic phase diagram that one expects in the presence of a 
glass transition, contains a pressure line that bifurcates from that 
of the liquid at some $\eta > \eta_{M}$ and which diverges at $\eta_{J}$ $[6]$. 
The bifurcation point is connected with the onset of glass transition.
The pressure of a system can be found from the knowledge of single and
two particle density distributions. For example, the virial pressure of 
a system in 3-dimensions is given as
\begin{equation}
\frac{\beta P}{\rho_{0}}=1-\frac{1}{6k_{B}TV\rho_{0}}\int d\br_{1} 
\int d\br_{2} \rho(\br_{1}) \rho(\br_{2}) g(\br_{1}, \br_{2}) 
(\br \cdot \bigtriangledown u(r))
\end{equation}
where $\br=\br_{2}-\br_{1}$. For an amorphous structure of hard 
spheres this reduces to
\begin{equation}
\frac{\beta P}{\rho_{0}}=1+\frac{2}{3}\pi \rho_{0}g(1)+\frac{4\pi}{V}
g(1)\sum_{q\neq 0} {|\mu_{q}|}^{2}S_{a}(q) j_{0}(q) 
\end{equation}

Note that for isotropic fluid the third term of above equation is zero; 
the bifurcation of pressure line from that of normal fluid starts as 
soon as particles start getting localized. Localization of particles also 
leads to crossover from nonactivated to activated dynamics and considerable 
increase in relaxation time.

\section{Summary and Perspectives}

A free-energy functional that contains both the symmetry conserved part of the 
DPCF $c^{(0)}(r)$ and the symmetry broken part $c^{(b)}(\br_1, \br_2)$ has 
been used to investigate the freezing of a system of hard spheres into 
crystalline and amorphous structures. The values of $c^{(0)}(r)$ and its 
derivatives with respect to density $\rho_0$ as a function of distance 
$r$ have been found using integral equation theory comprising the OZ 
equation and a closure relation proposed by Roger and Young $[22]$. 
For $c^{(b)}(\br_1, \br_2)$ we used an expansion in ascending powers of 
order parameters. This expansion involves higher-body direct correlation 
functions of the isotropic phase which in turn was found from the density 
derivatives of $c^{(0)}(r)$. For this we used the ansatz $[24]$ embodied in 
Eqs.$(3.10)$ and $(3.11)$. The contribution made by the symmetry broken 
term to the free-energy at the freezing point (liquid-crystal transition point)
was found to be about $8\%$ of the symmetry conserving part. Though this 
contribution is small but, as shown in Table $1$, it improves the agreement 
between theoretical values of the freezing parameters and the values found
from simulations. This result and the results reported earlier $[10]$ for 
the power law fluids show that the contributions of the symmetry broken 
part of free-energy increases with the softness of the potential. This 
explains why the RY free-energy functional was found to give reasonably 
good description of freezing transition of hard spheres fluid but failed for 
potentials which have soft core and/or attractive tail.
These results also indicate that the theory described here can be used to 
describe the freezing transitions of all kind of potentials. 

We used the free-energy functional to investigate the question of having
metastable states in between the homogeneous liquid and the regular crystalline 
state. The value of site-site correlation function $g(R)$ which provides the 
structural description of the amorphous structure have been taken from molecular 
dynamics simulation of granular system subjected to a sequence of vertical tapes $[21]$. 
The system has been found to behave like a glass-forming system. The reason for our 
choosing this data is that they are available at $\eta<\eta_{J}$ $[21]$ and therefore 
can be directly used in the theory without using approximations such as scaling 
relation $(5.1)$. Using the data of $g(R)$ at $\eta=0.576$ and $0.596$ from 
Ref $[21]$ we examined the stability of amorphous structures with respect to 
homogeneous fluid. The minimum of free-energy found at $\alpha \simeq 7$ suggests 
that the structures are  stable compared to that of the fluid and corresponds to 
a density distribution of overlapping Gaussians centered around an amorphous lattice.
This kind of structure may be associated with deeply supercooled states with 
a heterogeneous density profile. The transition of the liquid into any of these 
states will be determined by considering the dynamics of fluctuations around 
these minima. The glassy minimum is separated from the homogeneous liquid 
minimum by an energy barrier which height increases with the density.

Among future applications it will be instructive to investigate 
the contribution made by the second term of Eq.$(3.7)$ to free-energy 
of different potentials, application of the theory to freezing transition 
in two dimensions, in particular to examine the melting through hexatic phase
which other density functional schemes have failed to show and to examine 
the possibility of calculating total pair correlation functions of 
frozen phases using the OZ equation.

{\bf{\underline{Acknowledgments}}}: We are thankful to Dr. Massimo Pica 
Ciamarra and Dr. A. Donev for providing the data of $g(R)$. One of us (S. L. S.) 
is thankful to the University Grants Commission for research fellowship.
\newpage
\vspace{1.8cm}
\begin{center}
\begin{table*}
\caption{Freezing parameters of a hard-sphere fluid derived from the various
density functional schemes. Here $L= \left (\frac{3}{\alpha}\right)^{1/2}
\left( \frac{3\eta_{s}}{2\pi} \right)^{1/3}$ 
is the Lindemann parameter, $\eta_{s}=\frac{\pi}{6}\rho_{s}\sigma^{3}$ 
and  $\eta_{l}=\frac{\pi}{6}\rho_{l}\sigma^{3}$. Average errors are given 
in the parentheses. MWDA stands for modified weighted density approximation and
RY DFT stands for Ramakrishnan-Yussouff density functional theory.} 
\label{Tab1}
\begin{ruledtabular}
\begin{tabular}{|c|c|c|c|c|}
\small
{}&$\eta_{s}$& $\eta_{l}$& $\Delta \eta^{*}$ & $L$  \\ \hline
{Present result} &$0.542(<1\%)$& $0.490(<1\%)$ &$0.106 (<2\%)$  &$0.09$ \\ \hline
{MWDA-static reference [27]} &$0.503 (8\%)$ &$0.452(8\%)$ &$0.115 (10\%)$ &$0.13$ \\ \hline
{MWDA [28]} &$0.548 (< 1\%)$ &$0.474 (4\%)$ &$0.156 (49\%)$ &$0.10$ \\ \hline
{RY DFT [29, 30]} &$0.60 (10\%)$ &$0.511 (3\%)$ &$0.174 (69\%)$ &0.06 \\ \hline
{Simulation [25,26]} &$0.545$ &$0.493$ &$0.104$ &$\sim 0.13$  \\
\end{tabular}
\end{ruledtabular}
\end{table*}
\end{center}
\begin{figure}[h]
\includegraphics[height=3.0in,width=4.0in]{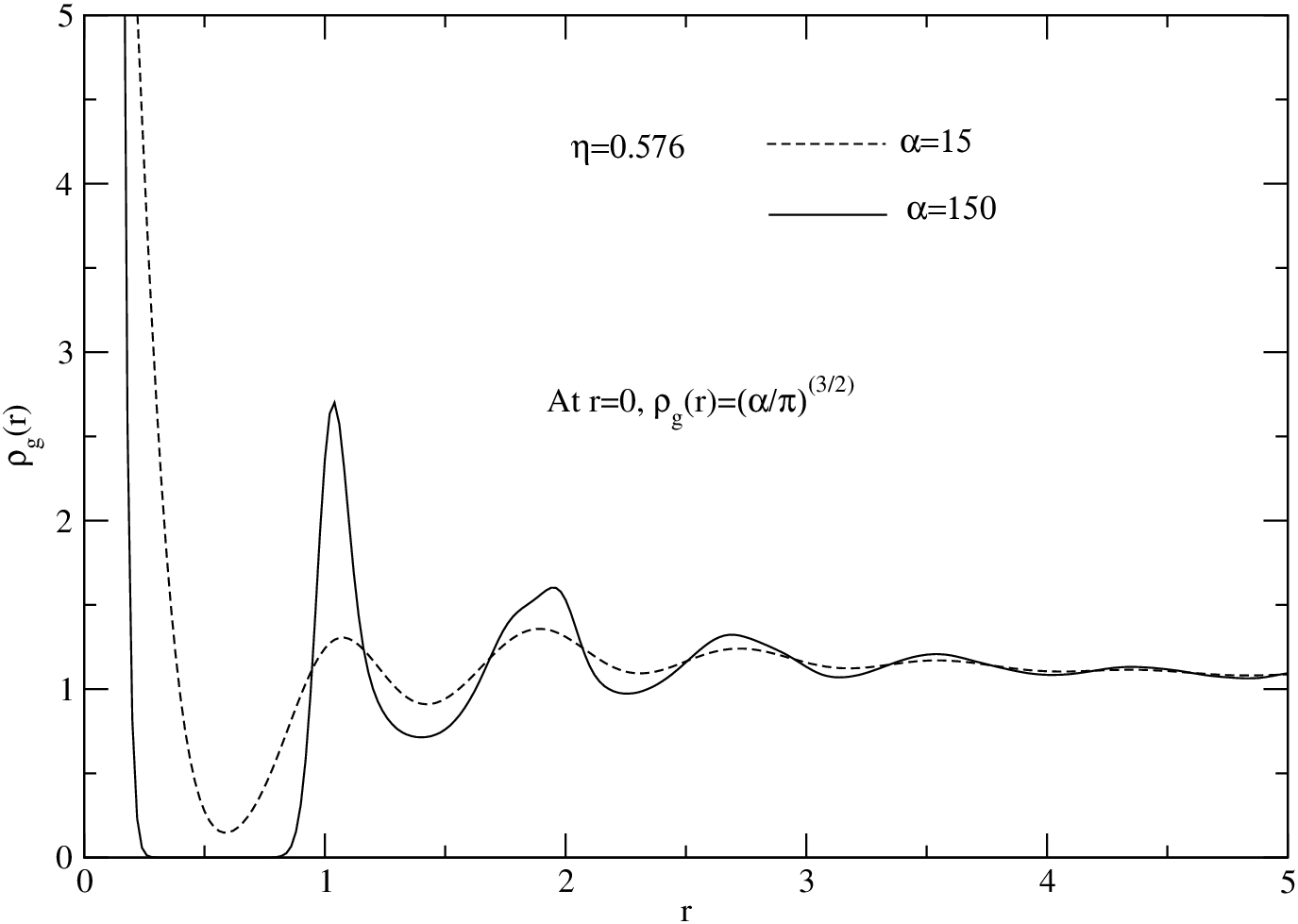}
\caption{Density $\rho_{g}(r)$ of an amorphous structure calculated from Eq.$(2.5)$
using the data of $g(R)$ found from molecular dynamic simulation of granular particles 
subjected to a sequence of vertical tapes for $\alpha=150$ (strong localization condition)
and $\alpha=15$ (weak localization condition)}
\end{figure}

\begin{figure}[h]
\includegraphics[height=3.0in,width=4.0in]{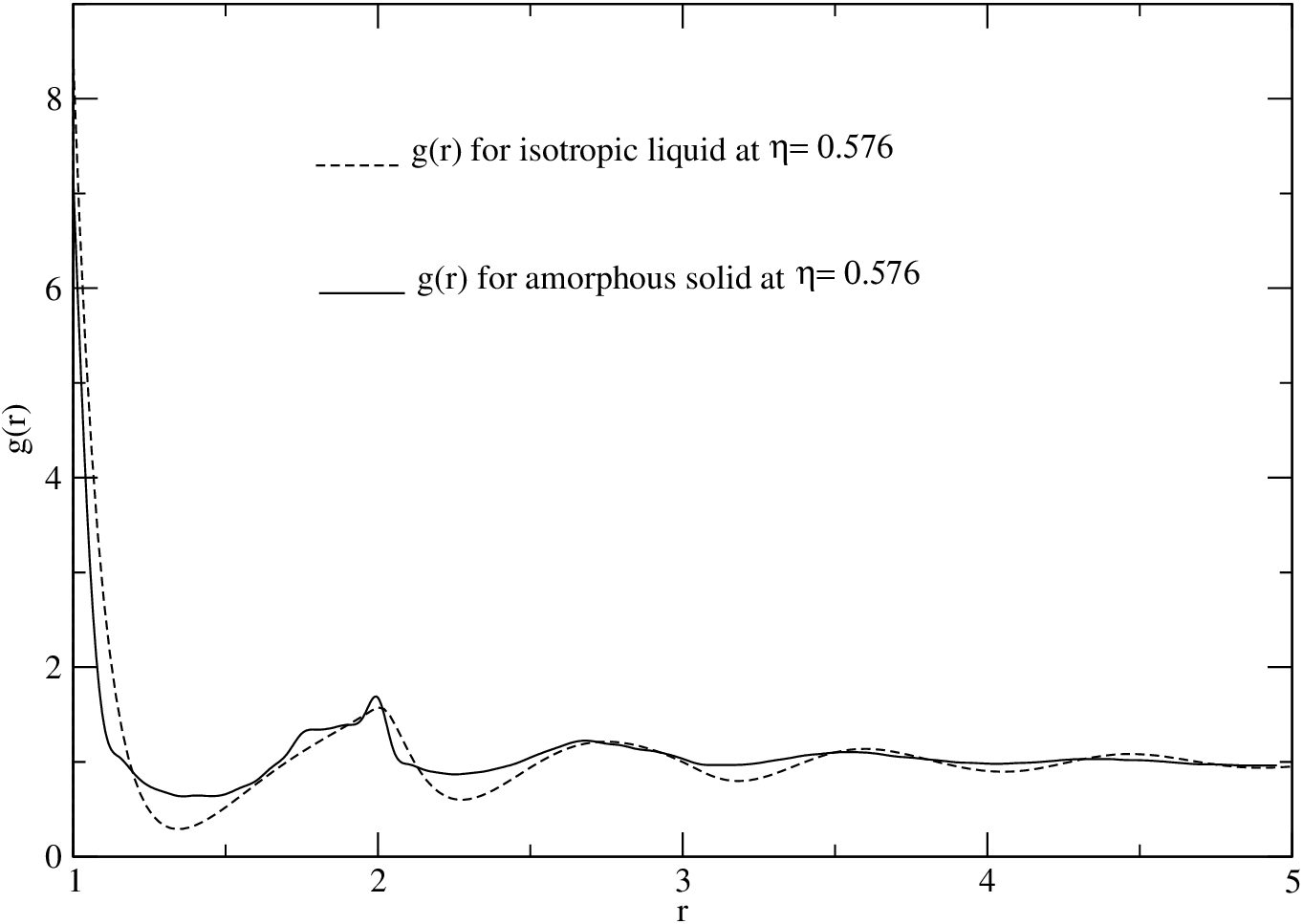}
\caption{Comparison of pair correlation function $g(R)$ of an amorphous structure and 
homogeneous liquid at the same packing fraction $\eta=0.576$}
\end{figure}

\begin{figure}[h]
\vspace{1.0cm}
\includegraphics[height=3.0in,width=4.0in]{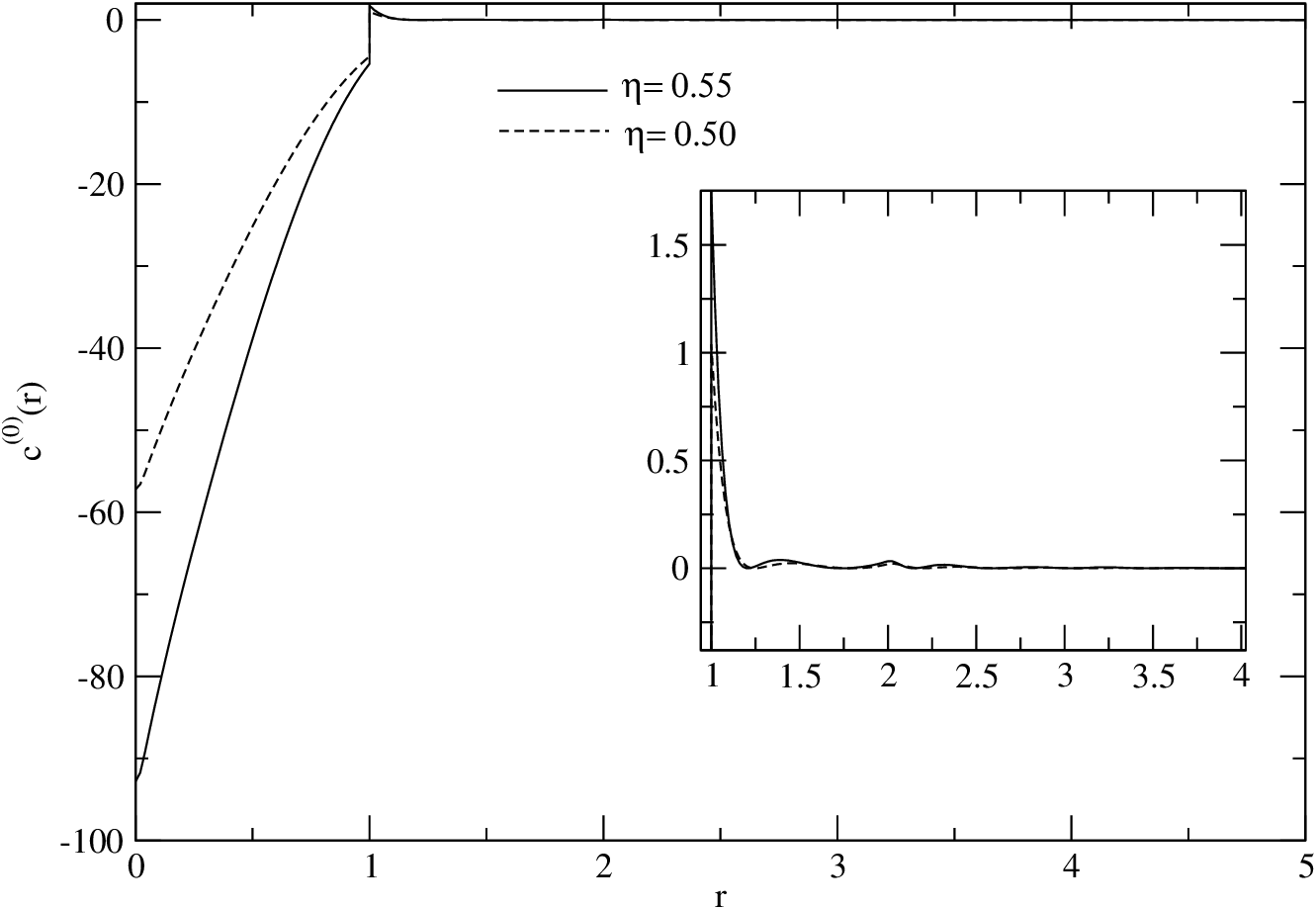}
\caption{Direct pair correlation function $c^{(0)}(r)$ as a function of distance r at 
packing fraction $\eta=0.50$ and $0.55$ found from the integral equation theory. The inset
shows at the magnified scale the value for $r\geq 1$}
\end{figure}

\begin{figure}[h]
\vspace{0.8cm}
\includegraphics[height=3.0in,width=4.0in]{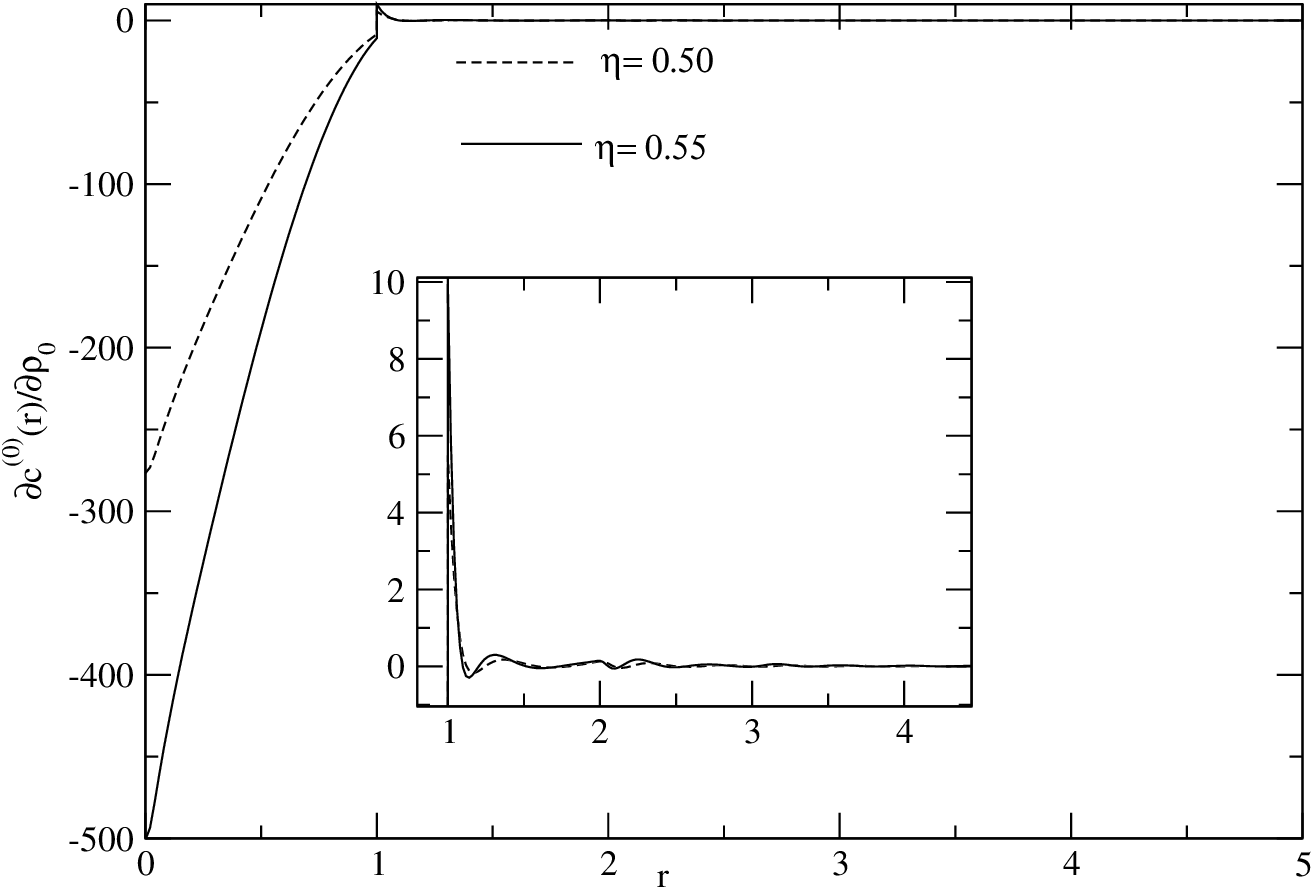}
\caption{Density derivatives of direct pair correlation function $c^{(0)}(r)$
at $\eta=0.50$ and $0.55$ found from the integral equation theory. The inset 
shows at magnified scale the value for $r\geq 1$.}
\end{figure}

\begin{figure}[h]
\vspace{0.8cm}
\includegraphics[height=3.0in,width=4.0in]{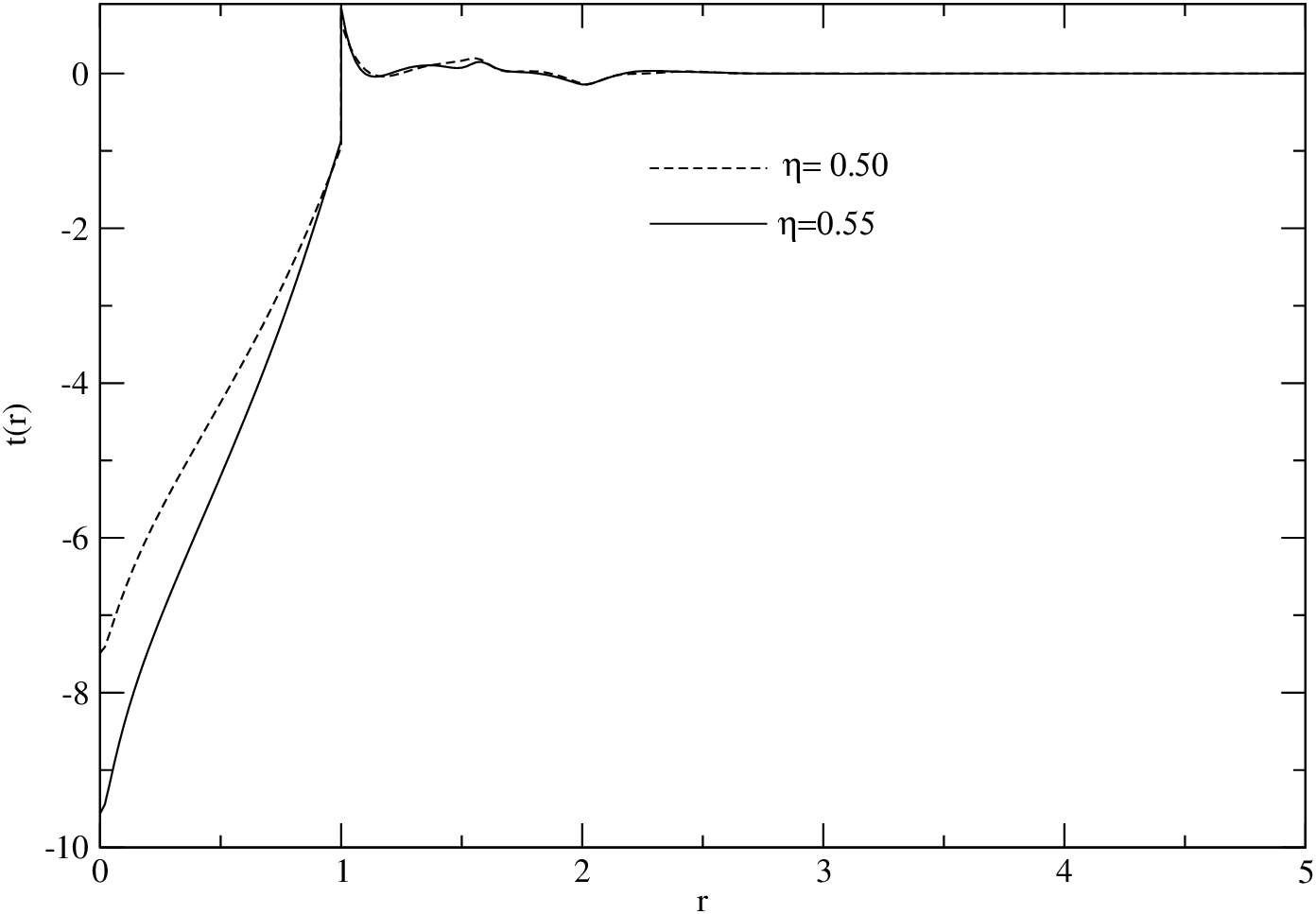}
\caption{Values of $t(r)$ as a function of distance for $\eta=0.50$ and $\eta=0.55$}
\end{figure}

\begin{figure}[h]
\vspace{0.8cm}
\includegraphics[height=3.0in,width=5.5in]{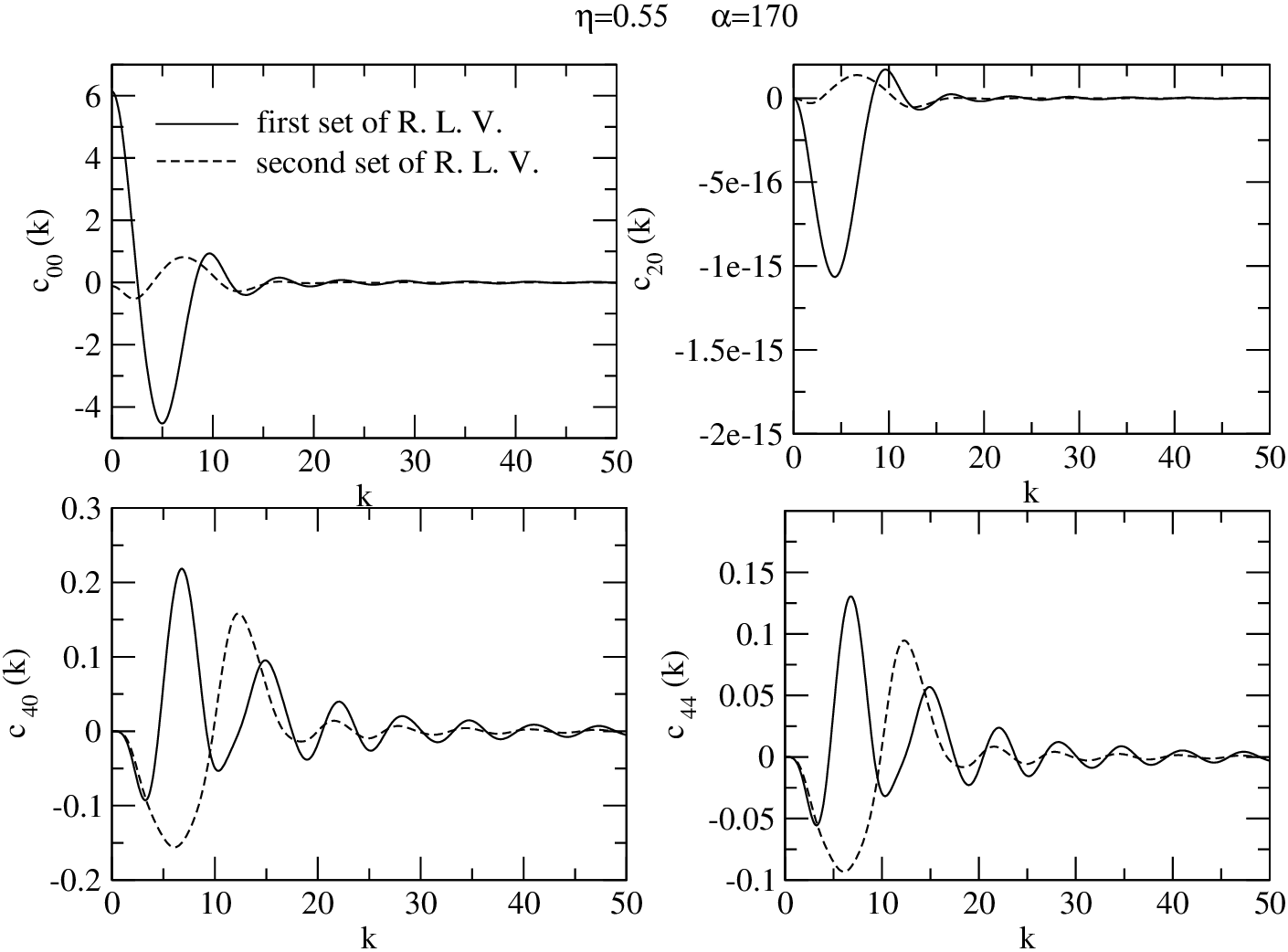}
\caption{Values of harmonic coefficients ${c^{(G)}}_{lm}(k)$ for 
first two sets of reciprocal lattice vectors of a f.c.c. crystal
for $\eta=0.55$ and $\alpha=170$}
\end{figure}

\begin{figure}[h]
\vspace{0.6cm}
\includegraphics[height=3.0in,width=4.0in]{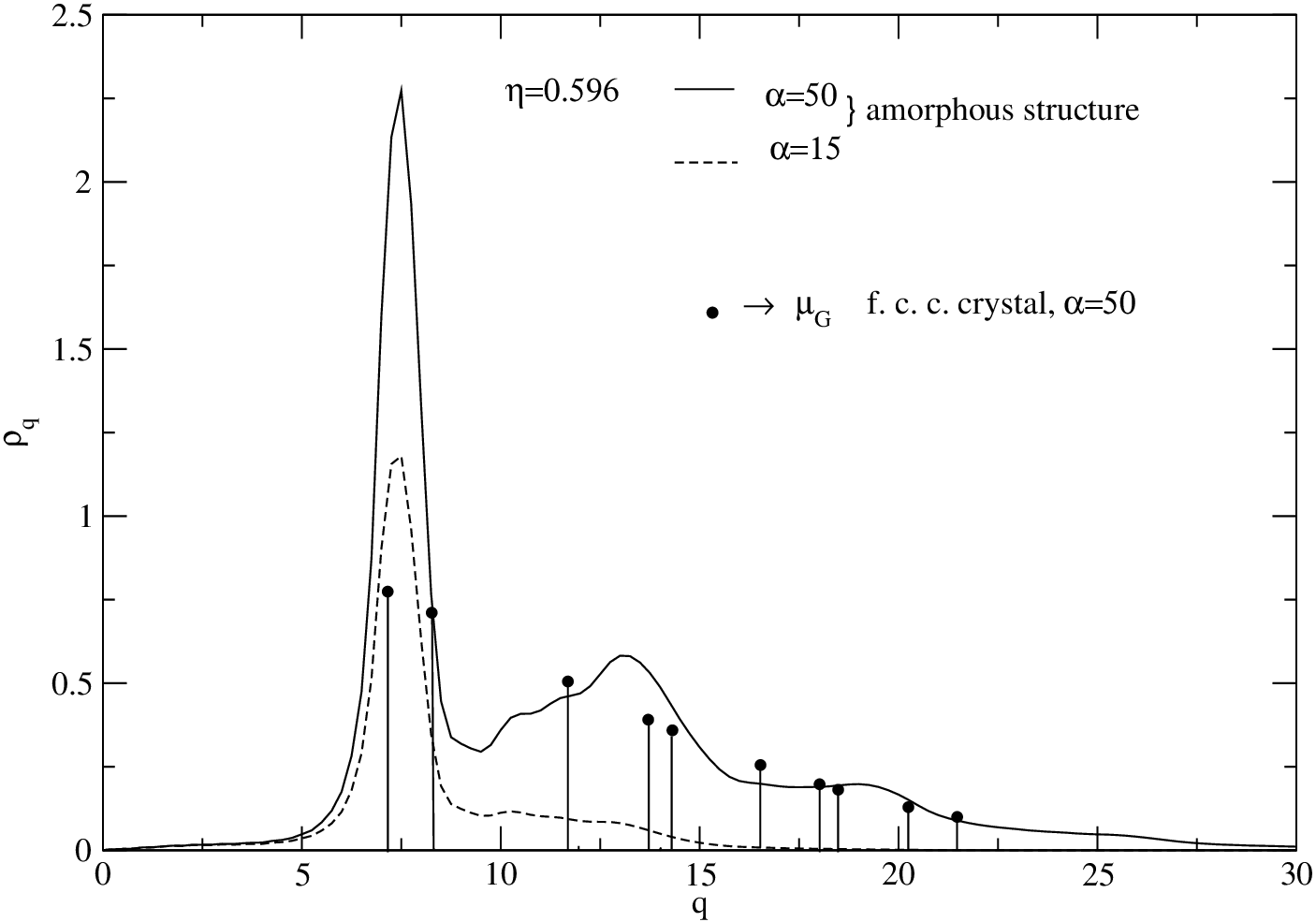}
\caption{Comparison of order parameters of amorphous structure for $\alpha=50$
and $15$ and of a f.c.c. crystal for $\alpha=50$ at $\eta=0.596$}
\end{figure}

\begin{figure}[h]
\vspace{0.8cm}
\includegraphics[height=3.0in,width=4.0in]{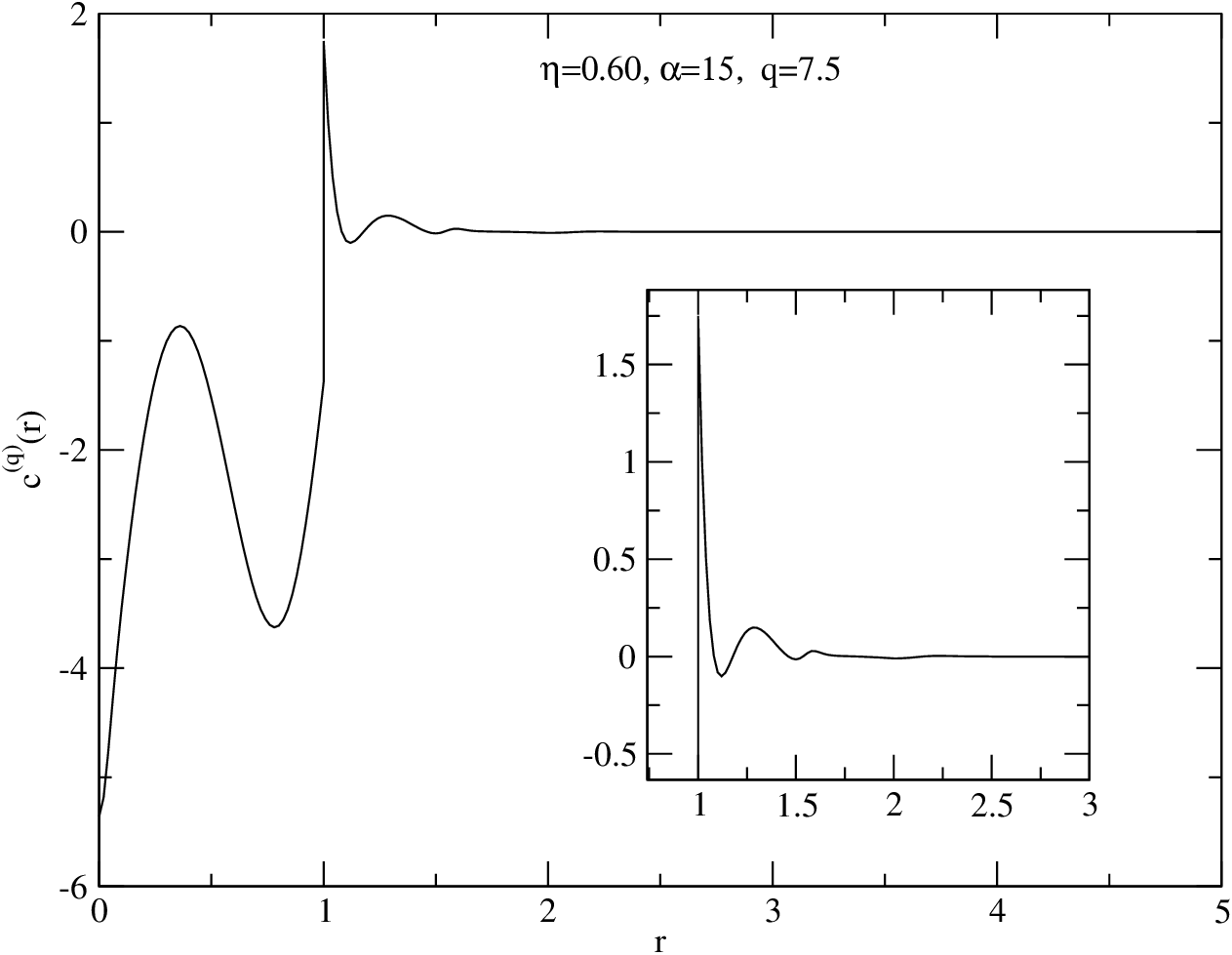}
\caption{Values of $c^{(q)}(r)$ as a function of distance $r$ at $q=7.53$
at which $\rho_{q}$ is maximum. Inset magnifies the value of $c^{(q)}(r)$
for $r\geq 1$}
\end{figure}

\begin{figure}[h]
\vspace{0.8cm}
\includegraphics[height=3.0in,width=4.0in]{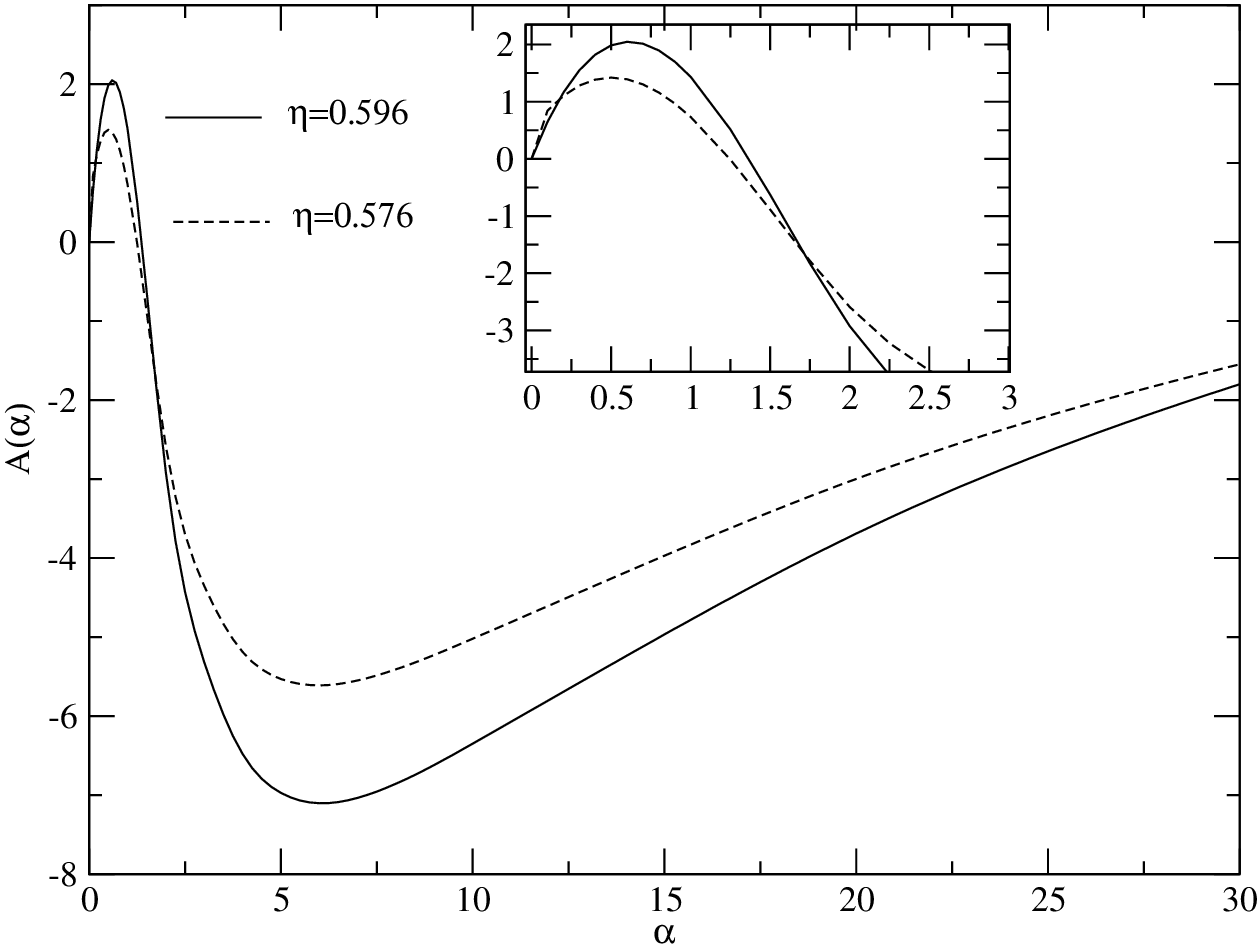}
\caption{Free energy difference $\Delta A=A[\rho_{0}]-A_{l}[\rho_{0}]$
as a function of localization parameter $\alpha$ at $\eta=0.576$ and $0.596$ 
for amorphous structures. The inset shows at magnified scale the energy barrier 
that separates the minimum of amorphous structure from that of homogeneous fluid.}
\end{figure}

\end{document}